\documentclass [acmsmall]{acmart}
\usepackage { graphicx }
\usepackage {makecell}
\usepackage{booktabs}
\usepackage{tabularx}

\AtBeginDocument{%
  }

\setcopyright{cc}
\setcctype{by-nd}
\acmJournal{PACMHCI}
\acmYear{2025} 
\acmVolume{9} 
\acmNumber{7} 
\acmArticle{CSCW355} 
\acmMonth{11} 
\acmDOI{10.1145/3757536}

\begin{document}

\title[I Want My Chart to Be Just for Me]{“I Want My Chart to Be Just for Me”: Community-Engaged Design to Support Outpatient Healthcare for Resettled Communities}

\author{Zhanming Chen}
\email{chen8475@umn.edu}
\orcid{0000-0002-9913-7239}
\affiliation{%
  \institution{University of Minnesota}
  \city{Minneapolis}
  \state{Minnesota}
  \country{USA}
}

\author{Juan F. Maestre}
\email{j.f.maestreavila@swansea.ac.uk}
\orcid{0000-0002-5403-9387}
\affiliation{%
  \institution{Swansea University}
  \city{Swansea}
  \country{United Kingdom}
}

\author{May Hang}
\email{may.hang@hennepin.us}
\orcid{0009-0006-8809-679X}
\affiliation{%
  \institution{NorthPoint Health and Wellness Center}
  \city{Minneapolis}
  \state{Minnesota}
  \country{USA}
}

\author{Alisha Ghaju}
\email{ghaju001@umn.edu}
\orcid{0009-0003-0870-4088}
\affiliation{%
  \institution{University of Minnesota}
  \city{Minneapolis}
  \state{Minnesota}
  \country{USA}
}

\author{Ji Youn Shin}
\email{shinjy@umn.edu}
\orcid{0000-0003-4978-3897}
\affiliation{%
  \institution{University of Minnesota}
  \city{Minneapolis}
  \state{Minnesota}
  \country{USA}
}
\renewcommand{\shortauthors}{Zhanming Chen et al.}

\begin{abstract}
Individuals resettled in a new environment often face challenges in accessing adequate healthcare services, particularly within the complex processes of outpatient clinic care. Cultural differences, language barriers, and low socioeconomic status contribute to these difficulties. While previous studies have identified barriers and proposed technology-mediated solutions for resettled populations, many focus on addressing deficits rather than building on the strengths these communities already possess, which limits the sustainability and relevance of these solutions in everyday life. We conducted two community-based participatory design workshops with 30 Hmong community members in a large metropolitan area in the US. Through this process, we identified four types of assets the community has gradually developed, including intergenerational support for health management and storytelling-based communication practices that facilitate relatable and culturally grounded interactions. We show how participatory design workshops can foster asset-based approaches, and discuss design implications for technologies that leverage patients' existing strengths to support their health management during outpatient visits. 
\end{abstract}



\received{October 2024}
\received[revised]{April 2025}
\received[accepted]{August 2025}

\maketitle

\section{Introduction}
Individuals may resettle in a new environment for various reasons, including economic opportunities, safety and security, or family reunification~\cite{cernea_RisksSafeguardsReconstruction_2000}. Those resettled from developing countries often encounter unique barriers across many aspects of life, including accessing timely, appropriate, and culturally responsive healthcare services. This can be due to factors such as differing cultural beliefs, precarious legal status, language barriers, and low socioeconomic status~\cite{kim_etal_ItsMyLanguage_2024, davis_randjelovic_CodesigningMobileDigital_2024, st-pierre_etal_RelationshipsPsychologicalDistress_2019}. Previous HCI and CSCW studies have examined factors that interfere with healthcare experiences in vulnerable populations, including children, women, and individuals from smaller cultural or ethnic groups. The findings indicate that critical factors for developing health technologies for these communities include language barriers, insufficient access to health insurance, and low health literacy and cultural competence~\cite{carballo_mboup_InternationalMigrationHealth_2005, chen_etal_DesigningHealthTechnologies_2025, mohammadi_etal_HealthCareNeeds_2016, ismail_kumar_EngagingSolidarityData_2018}. For example, one study used videos to educate pregnant women in the Global South about breastfeeding and birth preparedness~\cite{kumar_etal_ProjectingHealthCommunityled_2015}. Similarly, Tiwari and Sorathia suggested using photos and audio from medical providers to help teach low-literacy women to describe symptoms~\cite{tiwari_sorathia_VisualisingSystematizingPoor_2014}.

Asset-based community development (ABCD) is a well-researched approach to fortifying the existing capabilities of socially marginalized communities~\cite{mathie_cunningham_ClientsCitizensAssetbased_2003}. In contrast to needs-based methods, ABCD values each community’s capabilities, such as bilingualism. By driving changes through education, ABCD ensures that changes are not only sustainable but also more community-driven, leading to long-term benefits~\cite{karusala_etal_EngagingIdentityAssets_2019}. Following research in education and other social science fields, HCI and CSCW researchers have adopted ABCD and suggested new epistemological approaches to viewing marginalized communities as groups with their own knowledge and community wealth~\cite{wong-villacres_etal_CultureActionUnpacking_2020}. Researchers use ABCD as a lens to create opportunities for community members to reflect on and identify their own strengths~\cite{shin_etal_EveryCloudHas_2021, hardy_thebault-spieker_TurnAssetsCommunityBased_2024}. Various methods have been used to help visualize and mobilize these assets. Among them, community-engaged participatory design is particularly effective, as it equips end-users with skills and tools to identify their strengths and actively contribute to technology design from the early stages~\cite{wong-villacres_etal_ReflectionsAssetsbasedDesign_2021}. While asset-based approaches have benefits in the field, there is still a limited understanding of how communities in a new home gradually develop unique values, cultural strengths, and resilience over multiple generations. Furthermore, it remains underexplored how participatory design methods can support the identification of these assets and translate them into design directions reflecting the community’s lived experiences over time.

We conducted two community-based workshops with a small community in Minneapolis–Saint Paul, a large metropolitan area of the United States, focusing specifically on the growing Hmong-American community. With a history of resettlement and adaptation, this community provides valuable insight into the graduate development of assets, knowledge, and practices over time, in contexts such as the navigation of complex and technologically advanced healthcare settings. While this study focuses on the Hmong community, the insights gained can be applied to inform the design of healthcare technologies for individuals and groups who would benefit from building long-term resources and capabilities to support their health management. Our study seeks to understand the various assets the Hmong-American community draws on to manage their everyday health and navigate outpatient clinic visits. From these findings, we aim to inform the design of healthcare technologies for communities who have adapted and thrived in their new home.

Two main research questions (RQs) guided the study: (RQ1) What practices does a resettled community use to develop and sustain cultural strengths and values in managing everyday health, and what design opportunities can support these practices? (RQ2) How might we apply the lens of ABCD in designing technologies to leverage the community's unique strengths? This allowed us to effectively capture the lived experiences of community members, foster active conversations around their healthcare experiences, and invite them to envision community-centered healthcare technologies. Consequently, our contributions are threefold:

\begin{itemize}
  \item We show the usefulness of ABCD for HCI/CSCW in discovering unique strengths that a community built through adaptation and connection, over many generations.
  \item We highlight the significant role of culturally tailored community-based participatory design approaches in HCI/CSCW, stemming from the community's characteristics.
  \item We discuss how the strengths and specific needs of a resettled community serve as design implications of health technology.
\end{itemize}

In the following sections, we present related studies in HCI/CSCW, methods, findings, and discussion. 

\section{Related Work}

In this section, we cover key HCI, CSCW, and healthcare literature that examined the unique situational challenges faced by communities who have relocated to new environments, as well as approaches to support them through family- and community-centered technologies. We also explore how technologies could meaningfully support cultural practices, such as spiritual activities, during moments of challenge. In addition, we review ABCD as a framework for identifying and leveraging community assets, highlighting how participatory design methods could serve as tools and practices to mitigate the potential misrepresentation or neglect of those assets. Bringing together these perspectives, we emphasize the critical role of ABCD, combined with appropriate participatory design approaches, in exploring and supporting communities’ health management practices.

\subsection{HCI and CSCW Studies Supporting Health Management in Communities Adapting to a New Environment}

Communities living in new environments often face many challenges~\cite{internationalorganizationformigration_Migration_2024}. As they adjust to a new region, individuals experience barriers such as low socioeconomic status, language difficulties~\cite{gao_etal_TakingLanguageDetour_2022}, differences in cultural practices~\cite{morales_zhou_HealthPracticesImmigrant_2015}, and lack of resources~\cite{st-pierre_etal_RelationshipsPsychologicalDistress_2019}. These challenges often lead to marginalization across multiple aspects of life, including education, employment, and healthcare~\cite{baah_etal_MarginalizationConceptualizingPatient_2019, bascom_etal_DesigningCommunicationFeedback_2024}. HCI and CSCW studies have extensively explored these challenges, primarily focusing on infrastructural barriers to technology use in health management. Researchers have examined issues such as limited access to high-speed internet~\cite{bacishoga_etal_RoleMobilePhones_2016}, low technology literacy~\cite{seguin_etal_CodesigningDigitalPlatforms_2022}, and regulatory constraints imposed by immigration authorities~\cite{sabie_etal_DecadeInternationalMigration_2022}, and have proposed strategies to improve technology adoption and access for marginalized populations. However, addressing infrastructural barriers alone does not capture the full complexity of health management in marginalized communities, where everyday care practices are often shaped by family support and cultural resources. Building on this perspective, we focus on two main lines of literature: (1) technologies designed from community- and family-centered perspectives, and (2) technologies that leverage cultural practices to support health management within these communities.

Studies have examined how families navigating resettlement complement their language skills and technology literacy via intergenerational collaboration. In particular, given that second-generation youth typically has greater bilingualism and technology fluency, prior work has examined how children often serve as information mediators within families~\cite{yip_etal_YouthInvisibleWork_2022, le_etal_FeltWasDoing_2024}. In their role as mediators, children will search for information for various types of information, including education, healthcare, and everyday logistics (e.g., store hours or local regulations)~\cite{yip_etal_YouthInvisibleWork_2022}. For example, Le et al. highlighted how collaborative information search is approached by families adjusting to a new environment, and how these practices change over time in response to evolving family dynamics and situational needs, such as when children grow up and leave home~\cite{le_etal_FeltWasDoing_2024}. Similarly, studies have applied community-centered perspectives to design culturally tailored healthcare technologies that address the specific needs of resettled communities~\cite{buono_etal_MultimediaTechnologiesSupport_2019, seguin_etal_CodesigningDigitalPlatforms_2022}. These studies highlight the importance of access to social support and fostering collaborative health management among community members---an especially valuable approach in communities adjusting in a new environment, where social connections can play a critical role in helping individuals navigate unfamiliar healthcare systems, access information, and overcome barriers such as language and cultural differences~\cite{brown_grinter_DesigningTransientUse_2016, claisse_etal_UnderstandingAntenatalCare_2024}. For example, Tachtler et al. applied a socioecological approach to examine barriers to accessing mental health interventions for unaccompanied young people adjusting to a new environment, expanding the focus to include socioeconomic, political, and temporal factors~\cite{tachtler_etal_UnaccompaniedMigrantYouth_2021}. They suggested that, in addition to addressing daily life challenges, designs should account for the adaptability of interventions over time, as well as safety and privacy considerations when sharing sensitive mental health information (e.g., psychological status). Similarly, Ayobi et al. recommended fostering community connections through supportive online and offline activities as a useful strategy for designing mental health applications for resettled women~\cite{ayobi_etal_DigitalMentalHealth_2022}.

Another group of studies explored how cultural practices, including spiritual/religious practices and traditional remedies, offer accessible and trusted alternatives that reflect a community’s values and lived experiences~\cite{buie_blythe_SpiritualityTheresApp_2013, ibrahim_etal_TrackingRamadanExamining_2024}. This is particularly important in contexts where formal healthcare is unfamiliar, inaccessible, or misaligned with individuals’ worldviews or expectations. In such situations, people often turn to their cultural practices for comfort, trust, and support~\cite{wyche_etal_SacredImageryTechnospiritual_2009, wyche_etal_TechnologySpiritualFormation_2006, smith_etal_ThoughtsPrayers_2023}. In their literature review, Buie et al. introduced the term techno-spirituality to describe communication technologies that assist with individuals’ spiritual practices~\cite{buie_blythe_SpiritualityTheresApp_2013}. This study discussed how spiritual technologies can offer personalized, trusted, and culturally meaningful alternatives to formal systems. Claisse and Durrant extended the discussion of techno-spirituality by examining how faith-based practices were mediated through technology, specifically focusing on members of a Buddhist community in the UK during the COVID-19 pandemic~\cite{claisse_durrant_KeepingOurFaith_2023}. They found that online platforms became integrated into the community practice of Buddhism, enabling members to maintain connection and foster comfort by sharing their faith despite physical isolation. Similarly, Smith et al. applied spiritual support as a theoretical lens to understand social support within informal caregiving contexts~\cite{smith_etal_WhatSpiritualSupport_2021}. Their study showed that faith-based practices can provide emotional comfort and strengthen social connections in caregiving situations, where circumstances may be beyond an individual's control.

These studies commonly showed the potential of technologies to support spiritual and reflective practices during situational challenges, such as life-threatening health conditions, the pandemic, or living in unfamiliar environments. In these circumstances, where people's worldviews or expectations may be significantly disrupted, they often turn to their cultural and faith-based practices for comfort, trust, and support.

\subsection{Asset-Based Community Development in HCI and CSCW}

\subsubsection{Asset-Based Community Development as a Lens for Discovering Community Assets}

As a strategy for the sustainable community-driven approach, ABCD focuses on leveraging the community's unrecognized assets, capacities, and strengths in understanding and creating opportunities for the community~\cite{kretzmann_mcknight_BuildingCommunitiesOut_1993, mathie_cunningham_ClientsCitizensAssetbased_2003, yosso_WhoseCultureHas_2005}. It differs from the need-based approaches, which identify solutions based on communities’ needs, problems, and deficits~\cite{wong-villacres_etal_NeedsStrengthsOperationalizing_2020}. Because the needs-based approach emphasizes what the community lacks, it primarily values external support. It often overlooks the existing resources and opportunities within communities, which are crucial for sustainable practices. ABCD, conversely, values these community resources, offering development built upon what communities already possess, including social capital and human resources~\cite{hardy_thebault-spieker_TurnAssetsCommunityBased_2024}. 

Recent HCI and CSCW studies have applied ABCD when designing technology in marginalized communities, by examining underutilized assets and applying them as useful technology design concepts. Researchers conducted studies with individuals newly adjusting to the environment~\cite{pei_nardi_WeDidIt_2019, xu_maitland_MobilizingAssetsDatadriven_2017, chen_etal_EmpoweringFarmingCommunities_2025}, people of color~\cite{harrington_etal_DeconstructingCommunitybasedCollaborative_2019}, individuals in the Global South~\cite{karusala_etal_EngagingIdentityAssets_2019, wong-villacres_etal_DesigningIntersections_2018}, low socioeconomic status communities~\cite{cho_etal_ComadreProjectAssetbased_2019, cao_etal_AILiteracyUnderserved_2025}, and caregivers in chronic illness contexts~\cite{shin_etal_EveryCloudHas_2021}. Moving beyond community deficits, these studies emphasized the potential of community assets in technology design.

A few studies particularly examined communities navigating resettlement, and identified their unique assets and strengths that can be leveraged for sustainable development in different contexts, such as education. For example, Wong-Villacrés et al. performed ethnographic fieldwork in Hispanic communities in the US, and suggested designing conversational agents to amplify the trust between parents with unique cultural backgrounds and teachers, which they identified as a community asset; this would also further enhance the academic performance of their children~\cite{wong-villacres_etal_CultureActionUnpacking_2020}. Another study utilized the close relationships between family members in a low-income Latino community, and designed an SMS system to disseminate local educational opportunities through personal connections~\cite{cho_etal_ComadreProjectAssetbased_2019}. By conducting interviews and observations in a resettled community, Irani et al. designed a mobile app that provides access to internal and external resources (e.g., educational and career) to facilitate and personalize the integration process~\cite{irani_etal_RefugeTechAssetsbased_2018}.

\subsubsection{Participatory Design Methods as a Tool to Identify and Build on Community Assets}

Asset-based approaches emphasize building on the existing strengths and resources of a community. In HCI and CSCW, participatory design complements asset-based perspectives by providing practical methods and techniques to engage community members as co-designers and active contributors to technology design~\cite{broadley_AdvancingAssetBasedPractice_2020, petterson_etal_PlayingPowerTools_2023}. Participatory design offers strategies for identifying, mobilizing, and amplifying community assets from the beginning of the design process. Its emphasis on collaboration, respect for local knowledge, and shared decision-making makes participatory design a valuable method for supporting and building upon the assets that communities already possess~\cite{simonsen_robertson_RoutledgeInternationalHandbook_2012}.

While participatory design is generally rooted in democratic and collaborative values; as well, in HCI and CSCW, it is applied in ways that focus on identifying problems and unmet needs that can be addressed through design solutions. This is partly because technology design in HCI has historically been shaped by a needs-based, problem-solving mindset---a topic widely discussed in the field~\cite{wong-villacres_etal_ReflectionsAssetsbasedDesign_2021, wong-villacres_etal_NeedsStrengthsOperationalizing_2020}. Such a needs-based orientation is especially common in contexts like healthcare, where the primary goal is to identify specific problems (e.g., materialized perspectives, informational needs, lack of patient-centered care) and address them through external, technology-mediated interventions designed and prescribed by highly skilled researchers and practitioners~\cite{harrington_etal_ExaminingIdentityVariable_2022}. Without careful attention to appropriate methods and thoughtful approaches, participatory design can easily shift toward needs-finding and problem-solving practices~\cite{harrington_etal_DeconstructingCommunitybasedCollaborative_2019, raman_french_ParticipatoryDesignSensitive_2022}. This tendency is particularly prominent when working with marginalized communities, where unexpected challenges---such as building trust or navigating power dynamics---may arise.

This challenge has motivated growing attention to the role of asset-based design research in HCI. For example, Wong-Villacrés et al. introduced two studies that implemented asset-based design endeavors with resettled community members and sex-trafficking survivors, to examine the premises of asset-based approaches in underserved contexts~\cite{wong-villacres_etal_ReflectionsAssetsbasedDesign_2021}. Their work highlights the importance of significant preparatory efforts and slow, incremental processes that enable reflection and action within the research practice. Similarly, Harrington et al. pointed out how design-centered approaches in HCI and CSCW can unintentionally shift the focus of participatory design---decentering the interests and knowledge of communities by emphasizing their deficits and problems rather than their existing assets~\cite{harrington_etal_DeconstructingCommunitybasedCollaborative_2019}. These studies demonstrate that the participation of community members alone does not guarantee the discovery of assets, and therefore requires careful and intentional efforts throughout the research process.

Building on these previous HCI and CSCW studies that have emphasized the importance of carefully adjusting participatory design approaches to support specific communities and contexts, our study offers reflections on ethical and methodological considerations when working with a cultural group that has developed distinct practices and values while living within the broader context of a developed country. We discuss how community assets can be carefully identified, respected, and translated into the design of practical healthcare technologies through participatory processes. Despite growing interest in asset-based approaches, little is known about how communities develop unique values, cultural strengths, and resilience over multiple generations in a new living environment. Furthermore, there is limited research exploring how technology design can support and sustain these community-driven practices in everyday life. Traditional practices and family-centered health management within communities resettled in a new environment have often been viewed as less legitimate, and are excluded from mainstream healthcare systems, which are largely shaped by Western medical models. In this study, we aim to develop new technology design implications grounded in what community members are already comfortable with and what they do well in their everyday health management, thereby positioning their existing practices as resources for design, rather than barriers to be overcome.

\section{Methods}
In this study, we carefully designed our research approach to implement asset-based principles within a participatory design workshop. This section presents how asset-based perspectives shaped the detailed processes of data collection and analysis, with particular attention to ethical execution of the research. Each step of the process was designed to elicit participants’ knowledge, skills, and everyday practices that could inform meaningful and sustainable design directions.

\subsection{Study Contexts} 

Our study involved the Hmong community in Minneapolis--Saint Paul, an urban area that has the second largest Hmong population in the US~\cite{im_FactsHmongUS_2025}. The Hmong, originally from Southeast Asia, have been migrating to the US since the mid-1970s. As of 2021, approximately 370,000 Hmong reside in the US~\cite{gerdner_HealthHealthCare_2024}, with household income levels and college graduation rates remaining below national averages. Despite their relocation, the Hmong community has preserved many of its cultural practices and lifestyle while adapting to the new environment, including their shamanistic practices and clan- and family-centered social structures. Shamanism is a dominant belief system in the community, with adherents considering souls and ancestral spirits integral to both daily life and health practices~\cite{lee_vang_BarriersCancerScreening_2010}. In many Asian countries, shamanic healing practices include the use of herbal medicine, elaborate rituals, spiritual recitations, protective amulets, and the strategic placement of auspicious objects~\cite{sultana_ahmed_WitchcraftHCIMorality_2019, cho_etal_ShamAInDesigningSuperior_2025}. Shamans often serve as mediators between the human and spiritual realms, and these spiritual practices are believed to possess healing powers that protect against illness and sustain harmony between people and spirits~\cite{lor_etal_WesternTraditionalHealers_2017}. Additionally, the Hmong community places a strong emphasis on family and clan. With many households being multi-generational, Hmong people cherish family cohesion, valuing respect for elders and family gatherings~\cite{xiong_etal_EngagingCulturallyInformed_2016}. These unique cultural characteristics offer a rich opportunity to explore how cultural practices and community strengths shape health management within the US healthcare system, where advanced medical technologies often coexist with barriers to care.

At the same time, navigating the US healthcare system has presented ongoing challenges for the community. Hmong Americans have faced health challenges, including a relatively high incidence of chronic illnesses, particularly diabetes and hepatitis B~\cite{sheikh_etal_PrevalenceHepatitisVirus_2011}. Several sociocultural factors influence these health outcomes, including limited English proficiency, limited understanding of preventive care, ignorance of Hmong culture, and a preference for traditional remedies~\cite{lee_vang_BarriersCancerScreening_2010}. Beyond these challenges, the Hmong community has shown resilience and drawn on strengths that have helped its members adjust to life in a new country. Mutual support, commitment to family and community, and the preservation of cultural knowledge have contributed to the community’s settlement and continued growth within American society~\cite{hirayama_hirayama_StressSocialSupports_1988, vang_etal_MentalHealthHmong_2021}. These strong community assets and unique cultural practices, alongside healthcare barriers, provide valuable opportunities to investigate the health management experiences of resettled communities and explore how ABCD can inform the community-engaged design of health technologies. While this study is grounded in the context of the Hmong community, the challenges and strengths explored here reflect broader patterns shared by many who navigate healthcare systems.

\subsection{Ethical Considerations}

Throughout the study, the primary goal of the research team was to represent the voice of the Hmong community, while minimizing not only the power imbalance between researchers and community members, but also the misunderstandings and misrepresentations of their genuine lived experiences~\cite{lee_etal_CollaborativeMapMaking_2017, lee_etal_StepsParticipatoryDesign_2017}. A lack of understanding of Hmong history, cultural background, and characteristics could limit the opportunity to hear the voices of the community members. As such, we carefully specified the study plan, including the composition of the research team and the structure and logistics of the workshop, such as the environment and planned activities. Our university’s Institutional Review Board provided ethical approval for this study.

\subsubsection{Positionality}

Authors brought prior experience working with various communities, including research with small farming communities, individuals with low income and/or low literacy levels, patients with chronic conditions, and their family caregivers. The foundation for the current study was deeply shaped by community-based research, and by close collaboration with each community. 

We have two Hmong community members on our study team. One (the third author) is a healthcare provider from the Hmong community who contributed to the study design, the development of data collection materials, and participant recruitment, and provided valuable insights during data analysis to support interpretation. The other is a student assistant (see the Acknowledgements section) who supported the development of culturally grounded data collection materials and helped facilitate the participatory design workshops. While the rest of the authors neither identify as Hmong nor have direct experience working with the Hmong population, our past practices of learning from individuals in contexts where assets might not be easily visible prepared the team to have greater sensitivity and understanding when working with Hmong community members. These experiences fostered deep community engagement throughout the research process and supported our ability to empathize with participants’ perspectives during workshop design and data interpretation.

During data collection and analysis, the team held regular meetings to discuss recruitment plans and collaboratively develop workshop materials to ensure cultural appropriateness and minimize potential harm. For example, our workshop involved scenario-based activities where participants discussed their opinions of three different scenarios about medical care (e.g., the use of Western medicine vs herbal medicine); these were based on anecdotes shared by community members as well as our previous study results, which reflect on patient–provider communication in the Hmong community~\cite{chen_etal_DesigningHealthTechnologies_2025}. We carefully reviewed the content with Hmong team members and their close networks to ensure that the scenarios were culturally sensitive and would not cause discomfort or emotional challenges. Similarly, during participant recruitment, the established relationships between our Hmong team members and the community members were valuable, allowing us to rely on existing trust and connections. These collaborative practices continued throughout the study, including data analysis, interpretation, and manuscript preparation.

\begin{figure*}
    \centering
    \includegraphics[width=.48\textwidth]{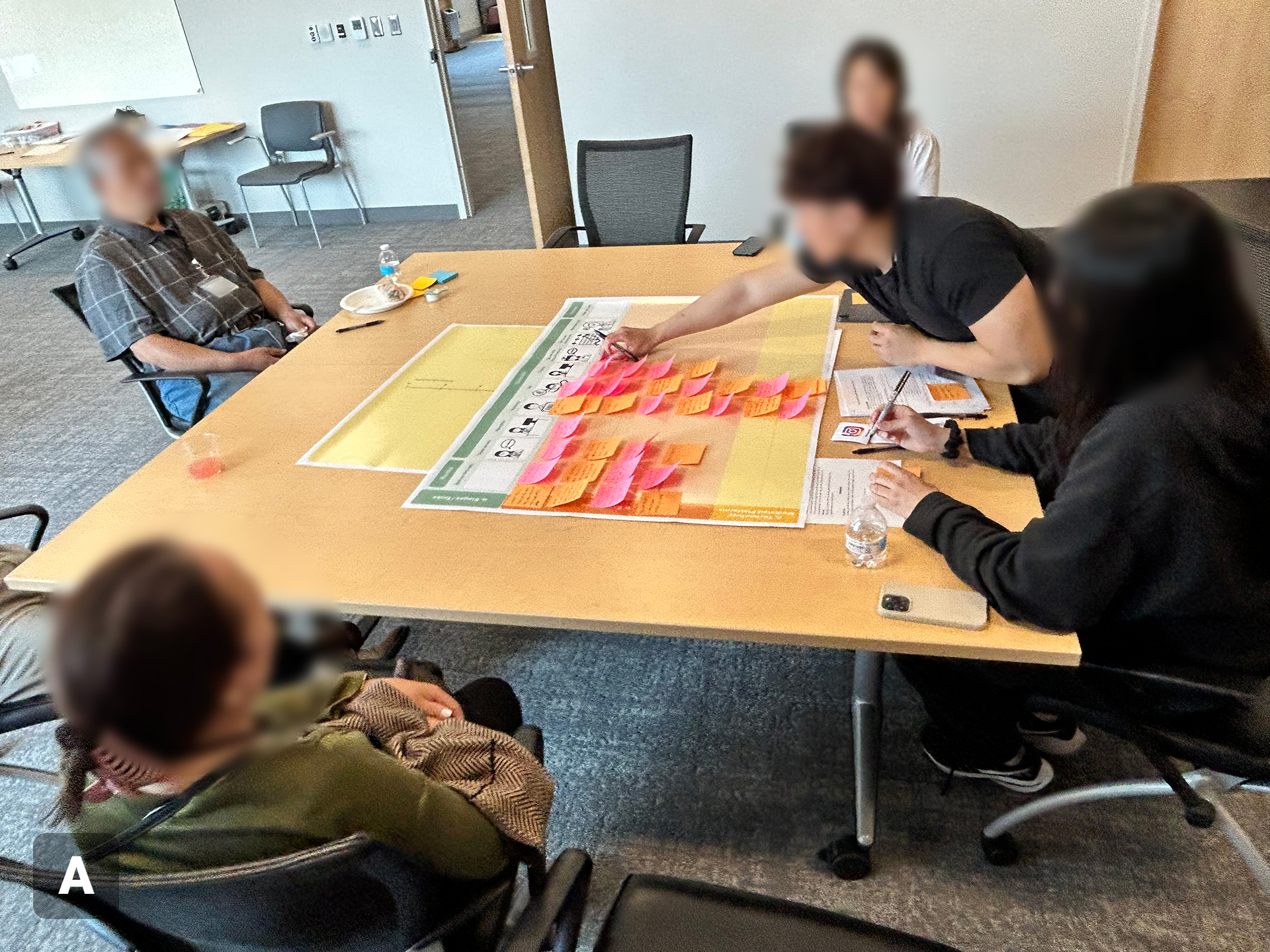}\hfill
    \includegraphics[width=.48\textwidth]{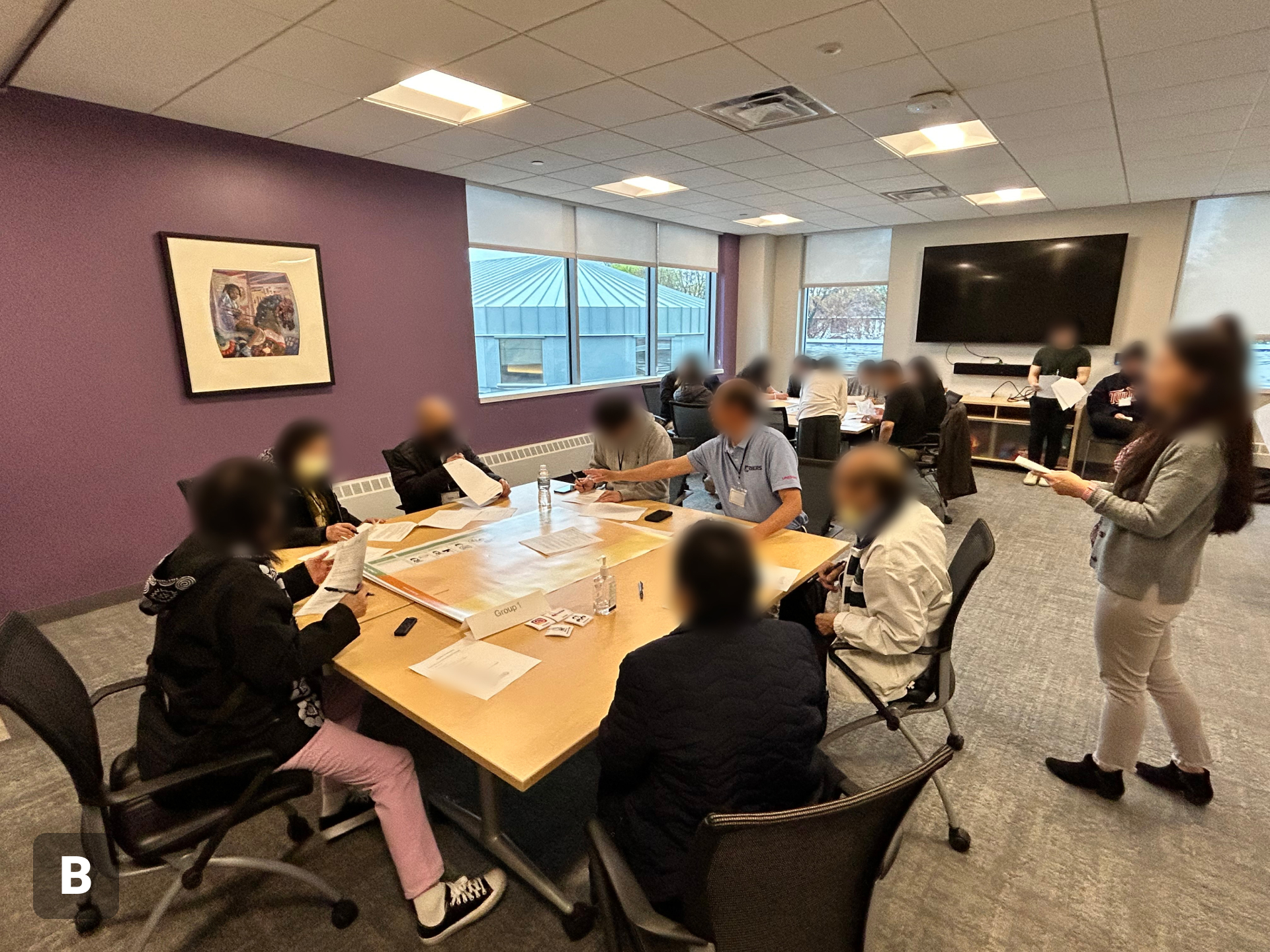}
    \caption{Community-based design workshops with participants. A: A facilitator arranged sticky notes with keywords from participants onto the user journey map. B: A facilitator provided a detailed explanation of the study overview and procedures to participants at the beginning of the workshop.}
    \Description{Community-based design workshops with participants. A: A facilitator arranged sticky notes with keywords from participants onto the user journey map. B: A facilitator provided a detailed explanation of the study overview and procedures to participants at the beginning of the workshop.}
    \label{fig:fig1}
\end{figure*}

\subsubsection{Workshop Design}

The workshop was carefully planned and structured to mitigate power imbalances between researchers and participants. At the beginning of the session, we introduced each team member by sharing their various connection to the community, and the aspects of their lived experiences that resonated with those of the participants. In particular, we introduced two Hmong team members, who supported the informed consent process, clarified demographic questionnaires, and facilitated participant engagement throughout the workshop. Similarly, we provided an overview of the study using everyday language rather than academic terminology. For example, the Hmong language lacks many formal or professional terms, such as ``research.'' Therefore, at the beginning of the workshop, we dedicated 30 minutes to carefully describing the study overview and procedures in detail (see Figure~\ref{fig:fig1}--B). Our goal was to ensure that Hmong community members clearly understood the purpose of the study and felt that their opinions and experiences were valued. 

To provide the most comfortable environment for participants, we prepared Hmong meals and dessert so that the participants could feel familiar. They were told they could have snack breaks anytime during the workshop; this also gave the researchers the chance to learn Hmong food traditions. Similarly, instead of having a large group discussion, we divided the participants into small groups, so that any introverts would have a greater chance to share their opinions. We deliberately assigned family members to different groups, to minimize issues of family dynamics (e.g., spouses relying on each other's opinions rather than expressing their own). 

Facilitators (research team members) were assigned to each group to guide the activities and support the participants. For example, if a facilitator noticed that an older participant had remained quiet for an extended period, they would gently encourage the participant to share by asking whether their experiences had been similar or different. To achieve more balanced participation within the group, facilitators sometimes invited participants to share a brief reflection, or even a single word, describing their experiences, and then asked them to expand on that in greater detail. During scenario-based activities, facilitators first encouraged participants to reflect on whether they or their family members had experienced those or similar situations, before inviting them to share potential solutions.


\begin{table*}
  \caption{Demographic breakdown of participants in two workshops}
  \label{tab:demog1}
  \setlength\tabcolsep{2pt}
  \begin{tabular*}{\textwidth}{@{\extracolsep{\fill}}p{5cm}cccc}
    \toprule
    & \textbf{Workshop 1} & \textbf{Workshop 2} & \textbf{Interpreters} & \textbf{Totals} \\
    \midrule
    \textbf{Age Range} & & & & \\
    \hspace{1em}21--30 & 3 & 0 & 2 & 5 (16.7\%) \\
    \hspace{1em}31--40 & 1 & 2 & 0 & 3 (10.0\%) \\
    \hspace{1em}41--50 & 3 & 2 & 1 & 6 (20.0\%) \\
    \hspace{1em}51--60 & 2 & 2 & 0 & 4 (13.3\%) \\
    \hspace{1em}Over 60 & 7 & 5 & 0 & 12 (40.0\%) \\
    \midrule
    \textbf{Gender} & & & & \\
    \hspace{1em}Female & 9 & 7 & 2 & 18 (60.0\%) \\
    \hspace{1em}Male & 7 & 4 & 1 & 12 (40.0\%) \\
    \midrule
    \textbf{Language Spoken} & & & & \\
    \hspace{1em}Hmong only & 8 & 6 & 0 & 14 (46.7\%) \\
    \hspace{1em}Hmong and English & 5 & 3 & 3 & 11 (36.7\%) \\
    \hspace{1em}Hmong and Lao & 1 & 2 & 0 & 3 (10.0\%) \\
    \hspace{1em}Hmong, English, Lao, and French & 1 & 0 & 0 & 1 (3.3\%) \\\
    \hspace{1em}English only & 1 & 0 & 0 & 1 (3.3\%) \\
    \midrule
    \textbf{Education} & & & & \\
    \hspace{1em}Some high school & 4 & 2 & 0 & 6 (20.0\%) \\
    \hspace{1em}High school grad & 0 & 4 & 0 & 4 (13.3\%) \\
    \hspace{1em}Some college & 6 & 3 & 1 & 10 (33.3\%) \\
    \hspace{1em}College graduate & 6 & 0 & 2 & 8 (26.7\%) \\
    \hspace{1em}Not applicable & 0 & 2 & 0 & 2 (6.7\%) \\
    \midrule
    \textbf{Annual Household Income} & & & & \\
    \hspace{1em}Less than \$10,000 & 0 & 1 & 0 & 1 (3.3\%) \\
    \hspace{1em}\$10,000--\$14,999 & 1 & 3 & 0 & 4 (13.3\%) \\
    \hspace{1em}\$15,000--\$24,999 & 2 & 0 & 0 & 2 (6.7\%) \\
    \hspace{1em}\$35,000--\$49,999 & 2 & 4 & 0 & 6 (20.0\%) \\
    \hspace{1em}\$50,000-\$74,999 & 1 & 0 & 0 & 1 (3.3\%) \\
    \hspace{1em}\$75,000--\$99,999 & 1 & 1 & 0 & 2 (6.7\%) \\
    \hspace{1em}\$100,000-\$200,000 & 2 & 0 & 0 & 2 (6.7\%) \\
    \hspace{1em}More than \$200,000 & 1 & 0 & 0 & 1 (3.3\%) \\
    \hspace{1em}Prefer not to answer & 6 & 2 & 0 & 11 (36.7\%) \\
    \bottomrule
  \end{tabular*}
\bigskip
\end{table*}


\subsection{Participants}
We recruited 30 participants from the Hmong community living in Minneapolis–Saint Paul, a large metropolitan area in the US. The first session had 19 participants, of whom 3 were fluently bilingual and served as interpreters; the second session included 14 participants, including those 3 bilingual members from the first session (see Table~\ref{tab:demog1}). The eligibility criteria for participants were being Hmong, being at least 18, having experience visiting healthcare institutions, and being willing to participate in the workshop activities. Our primary goal was to understand everyday healthcare experiences, rather than those pertaining to the management of specific diseases (e.g., chronic conditions). To achieve this, we recruited community members from an outpatient clinic at a local community healthcare institution, with the help of one of the authors, a healthcare provider who had gained the trust of the Hmong community and referred potential participants.

The median age of the participants was 52.5 years (ranging from 24 to 78 years), and 18/30 were female. 18/30 only spoke Hmong and/or Lao at home, while the other spoke both Hmong and English. Their annual household incomes ranged from less than \$10,000 (1/30, 3\%), \$10,000 -- \$14,999 (4/30, 13\%), \$15,000 -- \$24,999 (3/30, 10\%), \$35.000 -- \$49,999 (5/30, 17\%), \$50,000 -- \$74,999 (1/30, 3\%), \$75,000 -- \$99,999 (2/30, 7\%), \$100,000 -- \$200,000 (2/30, 7\%), to more than \$200,000 (1/30, 3\%). Most participants were married or in domestic partnership (20/30, 67\%), and most (20/30, 67\%) had not been to college.


\begin{figure*}
    \centering
    \begin{minipage}[b]{0.48\textwidth}
        \centering
        \includegraphics[width=\textwidth]{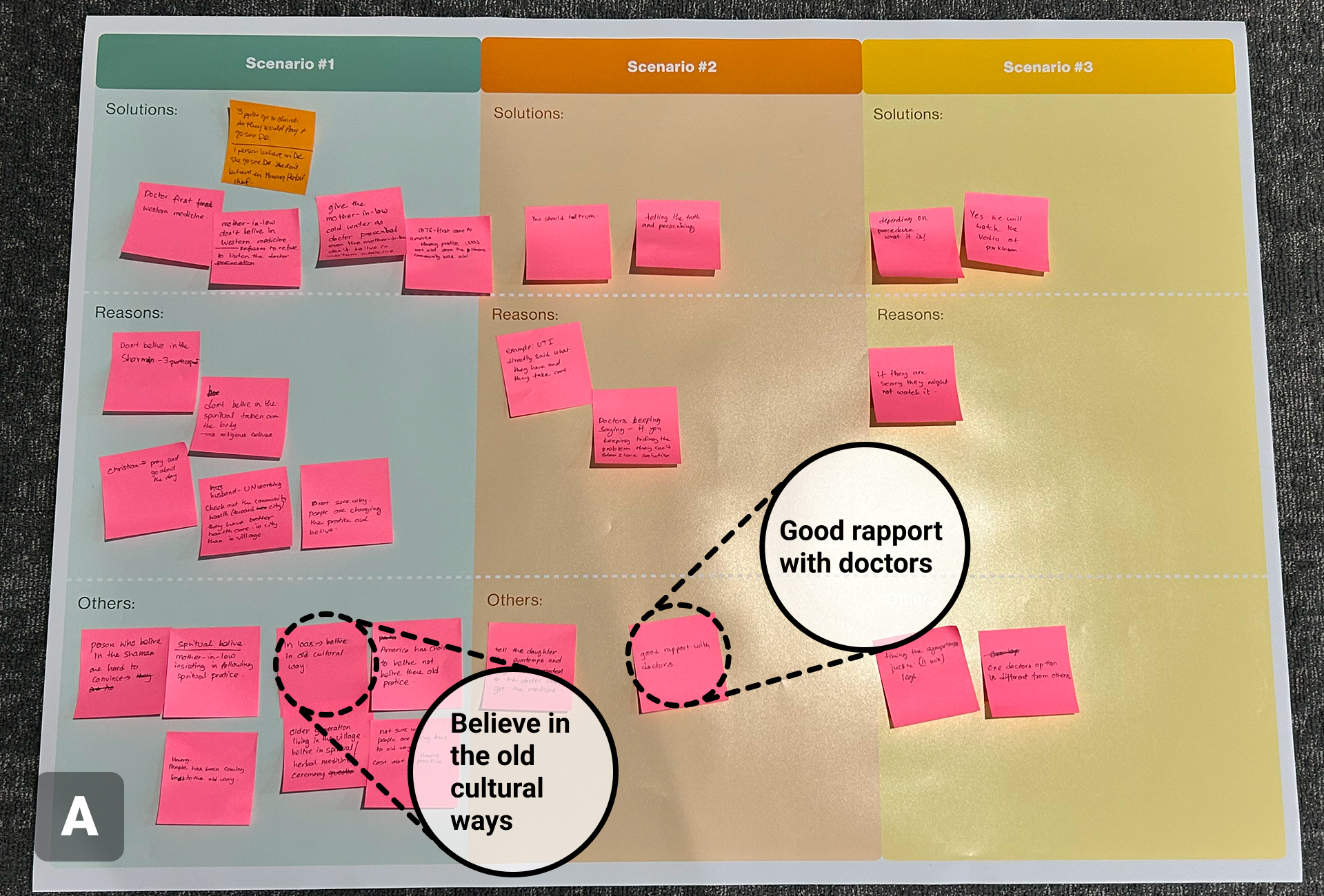}
    \end{minipage}\hfill
    \begin{minipage}[b]{0.48\textwidth}
        \centering
        \includegraphics[width=\textwidth]{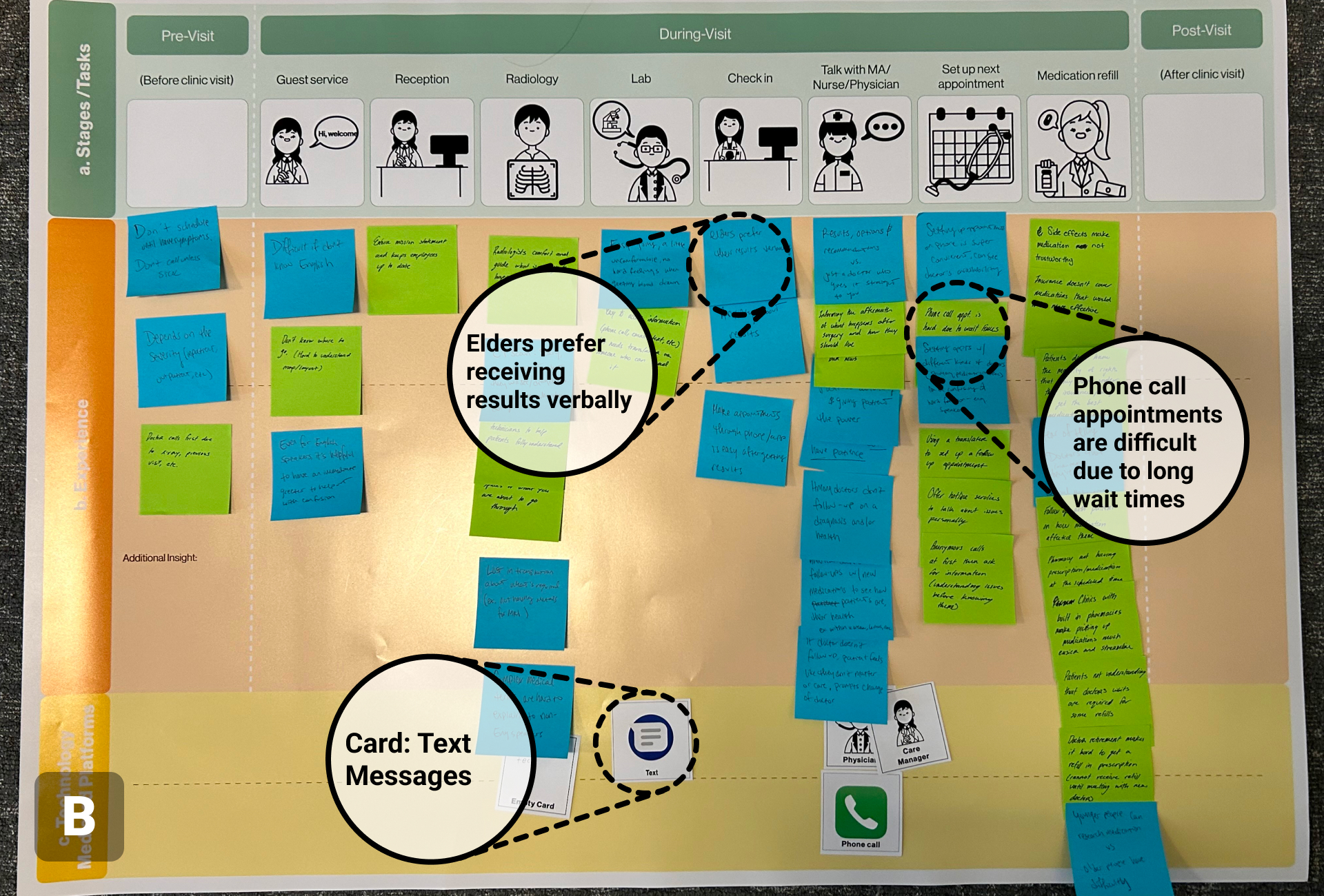}
    \end{minipage}

    \vspace{1em} 

    \begin{minipage}[b]{0.48\textwidth}
        \centering
        \includegraphics[width=\textwidth]{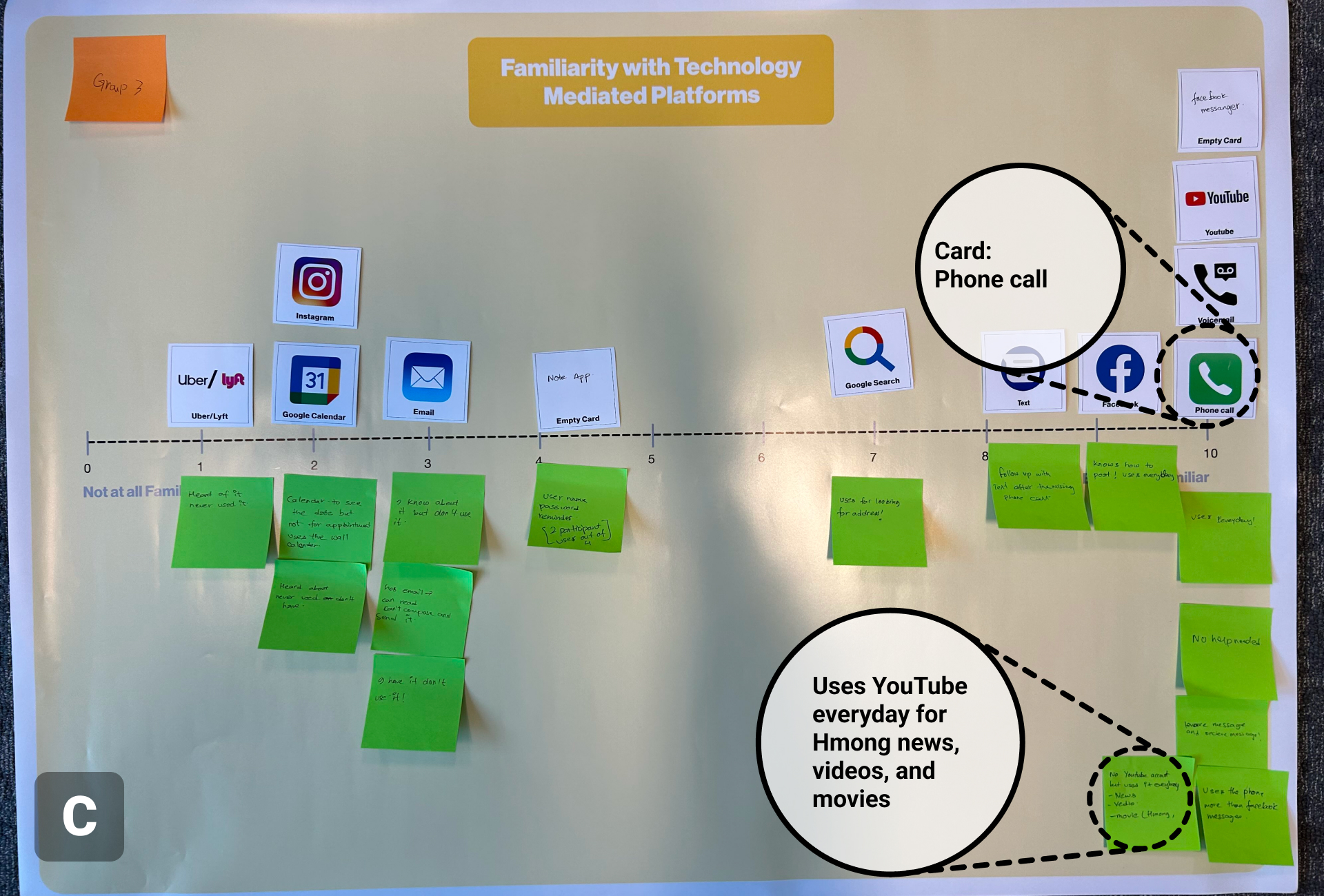}
    \end{minipage}

    \caption{Workshop materials. A: Scenario-based Discussion; B: User Journey Mapping; C: Technology Value Analysis.}
    \Description{Workshop materials. A: Scenario-based Discussion; B: User Journey Mapping; C: Technology Value Analysis.}
    \label{fig:fig2}
\end{figure*}


\begin{figure*}
    \centering
    \includegraphics[width=.48\textwidth]{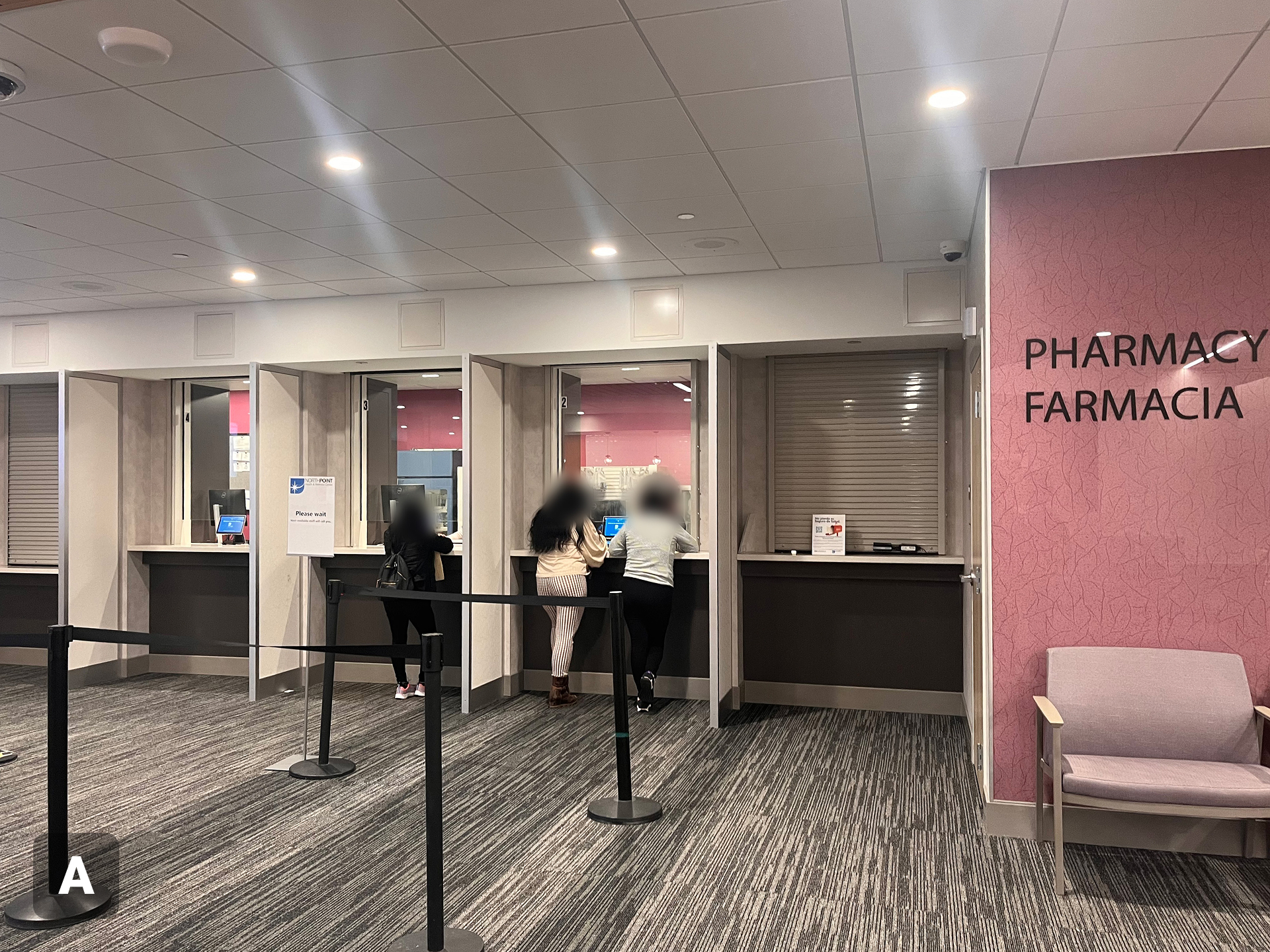}\hfill
    \includegraphics[width=.48\textwidth]{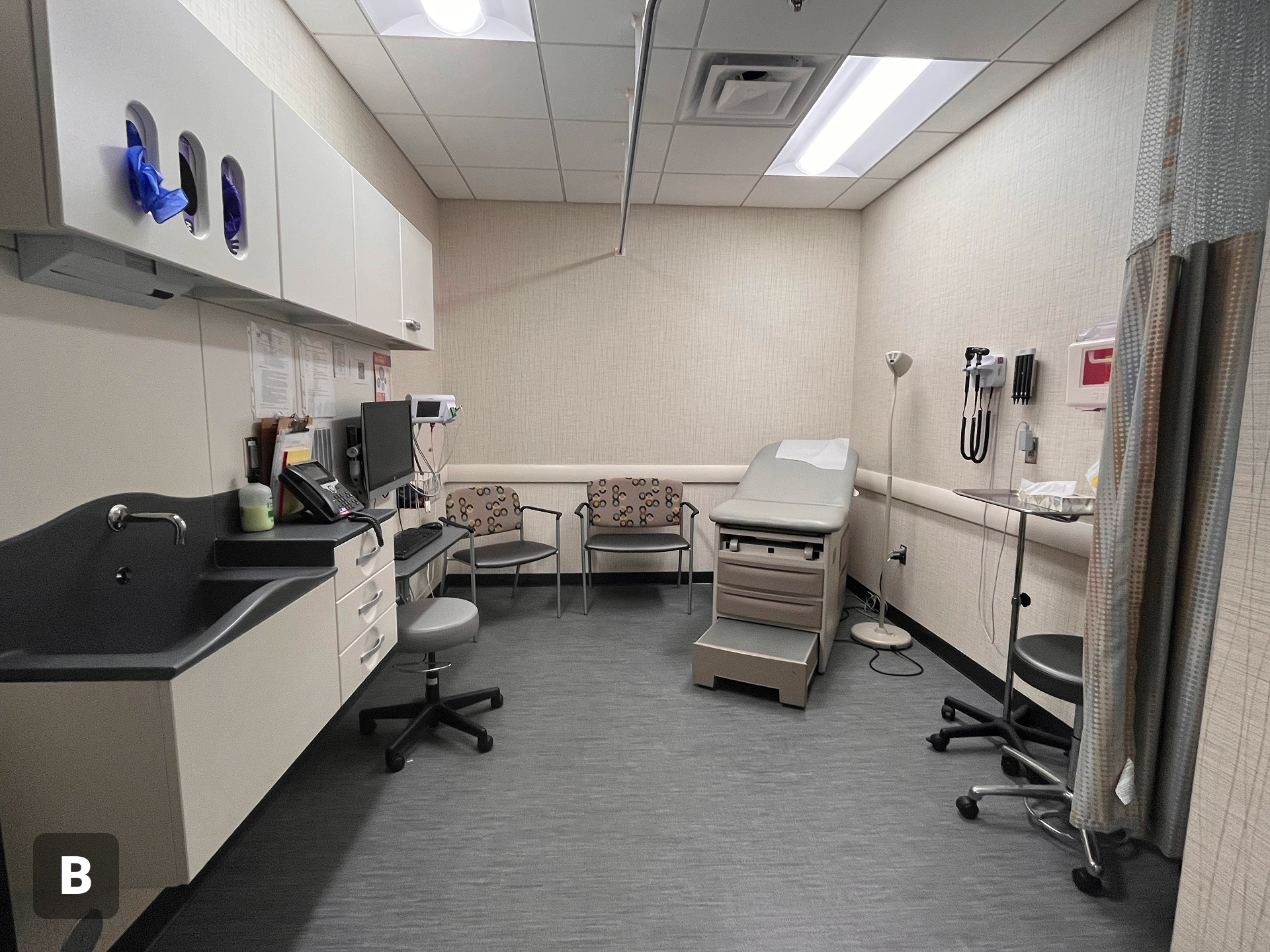}
    \caption{Clinic observation photos. A: The pharmacy of the clinic. B: The exam room of the clinic.}
    \Description{Clinic observation photos. A: The pharmacy of the clinic. B: The exam room of the clinic.}
    \label{fig:fig3}
\end{figure*}

\subsection{Activities Conducted During the Workshop} 

The workshop activities were carefully developed to reflect the community’s unique characteristics. To better understand these characteristics, we conducted site visits (observations) and semi-structured interviews with healthcare providers prior to the workshops~\cite{chen_etal_DesigningHealthTechnologies_2025, ghaju_etal_SupportingHealthcareProviders_2024}. These observations and interviews offered insights into the healthcare experiences of the Hmong community, as well as the clinical environment shaping patient care, including the clinic workflows, available resources, and types of providers involved. These insights informed our decision to use specific participatory design methods appropriate for the community (e.g., scenario-based discussions) and to develop the activity content (e.g., three health management stories) accordingly. Site visits took place in April 2024, during which we observed two community healthcare institutions that regularly serve Hmong patients (see Figure~\ref{fig:fig3}). One site we visited was an outpatient clinic where we later conducted our community-based workshops. The clinic serves a large number of underserved populations in the region, including the Hmong community, and has a longstanding commitment to addressing healthcare inequities linked to poverty. Conducting the workshops within this familiar and accessible setting helped increase community members’ participation and comfort levels~\cite{harrington_etal_DeconstructingCommunitybasedCollaborative_2019}.

A community member served as our guide and provided insights into the clinic’s layout and services, including the check-in process, informational forms outlining next steps for patients, and community-support services such as the food bank. With consent, we also shadowed a patient of color during their clinic visit. This observation offered additional insights into interpretation practices and the characteristics and preferences of Hmong patients. We also conducted a review of literature on community-centered health management, newcomers' health disparities, and Hmong communities. These practices led us to develop suitable and effective workshop materials and activities for the Hmong community.

Based on the insights from site visits and semi-structured interviews, we conducted two community-based design workshops in Spring 2024 (see Figure~\ref{fig:fig1}--A). Each workshop session lasted approximately three hours and included three main activities, which we describe below.

\subsubsection{Scenario-Based Discussion}
To elicit participants’ perspectives on critical health management issues, they were presented with third-person scenarios~\cite{kosow_gassner_MethodsFutureScenario_2008}. We designed three scenarios: shamanism as a treatment option; health information disclosure at healthcare institutions; and participants’ preferences and recommendations for health information materials that were common throughout their health management process. We then prompted participants to share their perspectives on each theme, reflect on similar experiences, and suggest possible coping strategies to alleviate the issue (see Figure~\ref{fig:fig2}--A). The scenarios were informed by the findings from the interviews, literature review, and discussion with two community members (see Table~\ref{tab:scena}).

\subsubsection{User Journey Mapping}
This activity was designed to understand the participants’ healthcare management experiences depending on how they interacted with healthcare providers and other stakeholders, which technology-mediated platforms they used, and which stage of the visit they were at (i.e., pre-visit, during-visit, and post-visit) (see Figure~\ref{fig:fig2}--B). In our clinic observations, we learned that a single appointment can involve visiting multiple sites, in addition to pre- and post-visit arrangements and follow-ups (see Figure~\ref{fig:fig3}). In addition to the pre- and post-visit phases, we further divided the during-visit phase into multiple sub-phases, including check-in and setting up the next appointment. 

\subsubsection{Technology Value Analysis}
The technology value analysis was designed to understand participants’ perceptions of widely used technology-mediated platforms, associated use cases, and barriers and opportunities for design interventions. Using small cards, the facilitators guided participants to take turns in speaking about their level of familiarity and comfort with technologies (see Figure~\ref{fig:fig2}--C).

\subsection{Data Analysis}
Each group's workshop discussion was audio-recorded for transcription and data analysis. Participants’ answers during the activities were handwritten on sticky notes; these were subsequently photographed and converted into digital data for qualitative coding. Although we had bilingual community members who translated Hmong-speaking participants’ answers during the workshop, the Hmong study team member further listened to the recordings and ensured the participants’ answers were properly translated and transcribed.

Following a constructivist approach~\cite{charmaz_ConstructingGroundedTheory_2014}, we conducted thematic analysis~\cite{braun_clarke_UsingThematicAnalysis_2006} to identify emerging themes regarding the Hmong community’s healthcare experiences. We entered the data collected from the workshops into NVivo. The first author performed an open coding for the transcripts and the content collected during the workshop activities. 


\begin{table*}
  \caption{Examples in the scenario-based discussion.}
  \label{tab:scena}
  \renewcommand{\arraystretch}{1.2}  
  \setlength{\tabcolsep}{6pt}  
  \begin{tabularx}{\textwidth}{%
    >{\raggedright\arraybackslash}p{0.05\textwidth}
    >{\raggedright\arraybackslash}X
    >{\raggedright\arraybackslash}p{0.34\textwidth}
  }
    \toprule
    \textbf{No\#} & \textbf{Main themes of the scenario} & \textbf{Questions} \\
    \midrule

    1 & This scenario highlights a situation where community members have both herbal remedies from a shaman and Western pills prescribed by their physicians as treatment options. & How would you behave in this case to develop the best plan? \\
    2 & This scenario describes a situation in which patients need to make choices about how much they want to disclose to caregivers and doctors. & What would you do to have effective treatment outcomes and feel comfortable? \\
    3 & This scenario describes options that patients have for using educational materials which explain clinical procedures and details of their care. & Would you be willing to use these materials to have an improved understanding of your care?   \\
    
  \bottomrule  
\end{tabularx}
\end{table*}


By applying ABCD approaches, we analyzed the data with a focus on recognizing the operationalized, practical tactics that participants use in their everyday health management practices. Rather than focusing solely on what participants know in principle, we centered our analysis on how they actually perform and navigate healthcare in practice---aiming to uncover overlooked assets and strengths embedded in their routines. We paid particular attention to: (1) patients' positive and negative experiences before, during, and after their clinic visits; (2) the unique characteristics of the community and their cultural practices; (3) community members' use of traditional treatments and Western medicine in varying situations; and (4) patients' use of technology in health management.

Our initial coding set included 396 low-level codes, which were iteratively refined and aggregated until they were mutually exclusive. This process resulted in 299 mid-level codes. During the coding process, we had a weekly group meeting to discuss the step-by-step progress. The first author presented the coding progress to the team during the meetings and discussed potential codes that can be added and aggregated based on their similarity in meaning to address any differences. We then conducted affinity clustering to generate emerging themes by grouping similar codes based on their relationships,  with a particular focus on the strengths and unique assets of the participants during their health management process. We identified and defined four high-level themes, each of which represents a key asset within Hmong community members' healthcare experiences: Balancing Cultural Healing Practices with Western Medicine as a Community Asset; Intergenerational Support as a Healthcare Asset When Privacy and Agency Are Respected; Social Ties as an Asset for Accessing Quality Healthcare; and Storytelling as a Cultural Asset in Healthcare for Relatable and Personalized Care. We present these findings in the following section. Direct quotes from participants were lightly edited for grammatical accuracy and clarity, without changing their intended meaning.

\section{Findings}

From two community-engaged workshops, we learned from our participants’ experiences that they have developed four unique assets---not merely carried over from their previous living environments, nor simply adopted from their new home, but shaped over time through lived experiences of resettlement, adaptation, and navigating between the two contexts. In this section, we present four types of strengths the community has built over time in managing health (see Table~\ref{tab:strate}).


\begin{table*}
  \caption{Four types of community assets in managing health}
  \label{tab:strate}
  \renewcommand{\arraystretch}{1.2}  
  \setlength{\tabcolsep}{8pt}       
  \begin{tabularx}{\textwidth}{%
    >{\raggedright\arraybackslash}p{0.24\textwidth}
    >{\raggedright\arraybackslash}X
  }     
    \toprule
    \textbf{Community Assets} & \textbf{Community Members’ Healthcare Practices} \\
    \midrule

    \textbf{Balancing Cultural Healing Practices with Western Medicine as a Community Asset}   
    & 
    \begin{minipage}[t]{\linewidth}
        \vspace{-2pt}
        \begin{itemize}
            \item A balanced approach between cultural practices and Western medicine 
            \item Viewing shamanistic practices as a means to provide comfort and respect for their culture 
            \item Involving community leaders (shamans) in collective clinical decisions 
            \item Informing healthcare providers about key cultural values
        \end{itemize}
        \vspace{5pt}
    \end{minipage} \\

    \midrule
    \textbf{Intergenerational Support as a Healthcare Asset When Privacy and Agency Are Respected} 
    &
    \begin{minipage}[t]{\linewidth}
        \vspace{-2pt}
        \begin{itemize}      
            \item Learning from wisdom and traditional values, including spiritual care and emotional support
            \item Younger family members assisting older adults with language and health technologies
            \item Resisting the use of certain healthcare technologies 
            \item Limiting family members’ involvement in certain aspects of care
        \end{itemize} 
        \vspace{5pt}
    \end{minipage} \\

    \midrule
    \textbf{Social Ties as an Asset for Accessing Quality Healthcare}
    &
    \begin{minipage}[t]{\linewidth}
        \vspace{-2pt}
        \begin{itemize}      
            \item Choosing appropriate providers based on recommendations
            \item Sharing healthcare plans among family and friends
            \item Building long-term relationships with providers
            \item Seeking healthcare and choosing providers of other ethnicities
            \item Avoiding the use of specific services for health management
        \end{itemize} 
        \vspace{5pt}
    \end{minipage} \\

    \midrule
    \textbf{Storytelling as a Cultural Asset in Healthcare for Relatable and Personalized Care} 
    &
    \begin{minipage}[t]{\linewidth}
        \vspace{-2pt}
        \begin{itemize}    
            \item Storytelling and indirect communication styles that made participants feel understood and respected 
            \item Conversational practices that are tailored to individual needs
            \item Follow-up phone calls as an expression of care and commitment
            \item Voice-based platforms to access and share health information
        \end{itemize} 
        \vspace{5pt}
    \end{minipage} \\

    \bottomrule
  \end{tabularx}
\end{table*}


\subsection{Balancing Cultural Healing Practices with Western Medicine as a Community Asset}

The first asset we identified was that community members have gradually learned the strengths of both traditional healing and doctors’ treatments, and they use each for different needs. During the workshops, participants shared how a balanced approach between cultural practices (e.g., herbal teas, crushed vines, and ointments prepared by family members) and Western medicine became a useful resource for managing health. Specific examples include using traditional medicine as a supplement, viewing shamanistic practices as a source of comfort, involving cultural leaders (shamans) in hospital decisions (e.g., treatment plans), and informing healthcare providers about key cultural values.

First, community members often use herbal medicine alongside Western medicine to manage symptoms. In Hmong culture, illness is understood to be influenced by diverse sources of power. While waiting for clinic appointments, community members often use herbal remedies to ease symptoms like pain or discomfort---especially when they believe the cause is related to spiritual imbalance or unwanted spiritual influences. One participant noted that older community members frequently combine prescribed medications with herbal treatments to reinforce a sense of holistic healing:

\begin{quote}
    \textit{For us, the older people, we search for help like the Hmong herbs and go to the counter in the pharmacy and buy some medicines. If these don’t work, I would say it is the ``shock'' that causes my illness. I massaged my hands at the point that caused the ``shock''. After doing this, you take the medicine, practicing shamanism, and everything will work. (Workshop 1, Group 1, P9)}
\end{quote}

In addition to using herbal medicine, community members often engage in spiritual practices such as prayer and consulting a shaman, alongside Western treatments, to reduce stress and maintain peace of mind. For community members, shamanism is a way to maintain emotional stability during uncertain situations, like treatments for long-term illness or injuries. P20 from the second workshop shared that praying is a way to maintain positive emotions rather than an attempt to directly treat their illness through practical means. This participant took medicine prescribed by the physician, and followed the recommendations for lifestyle changes, but still sought comfort through spiritual practices:

\begin{quote}
    \textit{I went to the doctor and took the medicine they gave me, and even though I took that medicine, I still asked God to help me and let the medicine heal me so I would be okay. My hope is for the doctor to heal me, but I also ask the sky or God to help me and let both work together so I can be well. (Workshop 2, Group 1, P20)}
\end{quote}

According to P20, community members emphasized the complementary roles of spiritual practices and Western medicine in the healing process. Their deeper understanding of the distinct strengths of both approaches motivated them to explain the value of their cultural traditions to providers, which enabled them to receive the care they desired.

During the workshop, participants also noted that providers sometimes appeared unfamiliar with certain cultural practices. They reflected that this unfamiliarity could lead patients to hesitate in sharing practices like shamanic healing, out of concern for being misunderstood or receiving negative reactions from providers.

\begin{quote}
    \textit{But then again, I could also see the other side where a Western physician might be kind of scared about the practices and stuff done and not really understanding. And that’s where having that bridge to connect the two might be the best pathway. (Workshop 1, Group 1, P7)}
\end{quote}

This participant always appreciates hearing stories about physicians who, although not Hmong, make an effort to learn about shamanistic practices—like coining, a traditional method that involves rubbing a coin on the skin to treat physical ailments—and how such practices affect the body. He shared that the providers’ openness, such as acknowledging the effects of coining practices, contributed to mutual respect and understanding between cultural practices and clinical treatments.

Despite efforts to provide learning opportunities for providers, participants pointed out that finding a balance between cultural practices and Western medicine is not easy. As a strategy, they have invited shamans and community leaders to help facilitate decisions about treatment options, including during clinical consultations. P1 suggested that deeper engagement between providers and cultural leaders in healthcare decision-making can create space for mutual respect and culturally responsive care:

\begin{quote}
    \textit{Well, I just want to say that I think we can provide space. I don’t think that Western medication and religious practices will always agree, but providing space for people from both sides to express themselves, and then providing space for patients to decide whether they want to bridge it or do their own thing at home. Sometimes patients want to have these conversations at the hospital. (Workshop 1, Group 1, P1)}
\end{quote}

For community members, seeing providers make an effort to understand their unique cultural practices in managing health creates positive impressions that support adherence to treatment.

During the workshop, we learned that not every community member follows cultural healing practices, especially among younger generations. Despite this, community members still value having choices. In the second workshop, P19 explained that even if a cultural practice does not have a direct medical effect, the positive emotions and sense of comfort it brings can have a synergistic effect when combined with a physician’s prescription:

\begin{quote}
    \textit{I believe in the Shaman. I believe it is doing its own work. But I do go to the doctor because I know that the doctor is better at diagnosing, and the other practices can add to the healing power. But in the end, it is really the doctor who heals you. (Workshop 2, Group 1, P19)}
\end{quote}

As such, many community members continue to use cultural practices as a source of emotional comfort and peace---an asset that should be recognized as part of a broader treatment approach.

This subsection highlights the community’s asset of balancing traditional healing practices with prescriptions from healthcare providers. To maximize this asset, community members have developed various approaches, including inviting community leaders and explaining cultural values to their providers. These efforts reflect the community’s flexibility and openness in integrating multiple forms of care to support both physical and emotional well-being.

\subsection{Intergenerational Support as a Healthcare Asset When Privacy and Agency Are Respected}

Intergenerational family collaboration is another asset that plays a significant role in shaping healthcare experiences within the Hmong community. Cultural values rooted in respect for elders, mutual care, and familial responsibility continue to guide how families navigate healthcare. While second- and third-generation family members are more familiar with Western healthcare systems and digital tools, first-generation elders often rely on traditional practices and understandings. These differences shape the unique dynamics of family-based health management.

During the workshops, participants described generational differences in familiarity with healthcare technologies, and in preferences for treatment approaches. These differences were often seen as opportunities for mutual support and learning. Their discussions showed how older generations offer wisdom and traditional values, while younger generations contribute skills in healthcare management.

Young participants in their 20s and 30s explained how they had learned from the wisdom and traditional knowledge of older family members and had gradually come to appreciate the cultural and spiritual traditions of the Hmong community. Although they tended to lean more toward Western practices and medications, over time, they began to reflect on their life, family, and community, and came to understand traditional practices as expressions of love, protection, and emotional solidarity within the family---an important asset that helped them thrive:

\begin{quote}
    \textit{I've come to understand the practices, the religion of my family members, whatever they may be, when they ask me to do something involving a shaman or a pastor or a prayer or something like that. Even though I'm more inclined to Western medication, I've come to understand as I've matured that they're asking me to do this, because they love and care for me. (Workshop 1, Group 1, P7)}
\end{quote}

P7 also shared how the presence of family and collective spiritual practices provided emotional strength and resilience during an intensive medical procedure, reinforcing familial bonds, support, and care. This participant mentioned that before a major surgery, their extended family traveled from other states to support and pray with them. At the time, they didn’t think much of it, but looking back, they realized the gathering was also important for the family’s reassurance and emotional support.

While the younger participants described gaining emotional support and cultural grounding from older generations, the older generations, in turn, often relied on younger family members---especially children and grandchildren---for support in communication at the clinics and hospitals. Due to limited English proficiency, older community members emphasized the value of having their younger family members as trusted and effective intermediaries during healthcare visits. P27 shared their preference for having a family member as their interpreter, highlighting the benefits of shared life knowledge, such as habits and the use of medications:

\begin{quote}
    \textit{If I could, I would prefer to have a family member interpret. They live with me, they know my habits at home, they know what kind of medicine I’m taking, and they understand what’s easy for me and what’s hard for me. So if I could have a family member as the interpreter, it would be less frustrating because I wouldn’t have to explain so much of the situation. (Workshop 2, Group 3, P27)}
\end{quote}

For P27, involving younger family members familiar with the English language, their healthcare practices, and the Western healthcare system helped reduce the efforts and burdens in explaining everything completely by herself. Younger generations are also seen as important sources of learning in the use of health management technology. Many elders reported difficulties using healthcare tools, such as the MyChart patient portal. Through support from younger relatives, older adults learned how to receive reminder messages from providers, manage online calendars, and communicate through patient portals. Participants shared how younger community members often introduced and taught these technologies to older adults in the family. P19 explained:

\begin{quote}
    \textit{As long as the providers think it’s time for me to go back for a test, they’ll call. And if they don’t call, they’ll send a letter or email my children. I use what they call MyChart---it’s on the computer. Then my children will check it. Maybe they’ll get a letter saying it’s time, and I can go in for a physical or a colonoscopy. They’ll take care of it for me. (Workshop 2, Group 1, P19)}
\end{quote}

This participant shared the experience of how their children help manage digital communications from providers, such as notifications and test results on MyChart. 

However, despite its benefits, intergenerational support in healthcare---particularly around technology use---also requires extra mindfulness of privacy. Older participants pointed out the importance of maintaining an adequate level of privacy in order to fully maximize the asset of intergenerational family support. Specifically, while they appreciated assistance with tasks like scheduling appointments, they did not always wish to share the full details of their health records. Our participants described two ways they maintained boundaries: resisting the use of certain healthcare technologies and limiting younger relatives’ involvement in specific aspects of their care. When their family members saw the after-visit summaries and lab results, they tended to interfere with the details and control over everyday health behaviors---such as what they eat or avoid---even though they only needed help with making appointments:

\begin{quote}
    \textit{I said, ``I don’t have any diabetes. Just a little bit of blood pressure.'' And [my daughter] said, ``I want to see it.'' Even if I let her see it, she’ll tell me what to eat and what not to eat. I can’t handle that. I eat what I want to eat. (Workshop 2, Group 2, P24)}
\end{quote}

As such, involving family members can interfere with their sense of agency, which they do not prefer. Feeling a lack of control led some older participants to intentionally limit family involvement in their use of digital health platforms. For example, using the patient portal often begins with a family member creating the account on their behalf. After that, managing information and navigating the platform typically requires ongoing family involvement. P5 shared that having a patient portal account means sharing every single detail of their health information with family members, including their son and daughter. This participant expressed discomfort, saying the portal should be for them---not the whole family:

\begin{quote}
    \textit{If I create a patient portal account, my son will know about it, and my daughter will know about it. And everybody will look at it... I don’t like that. I want MyChart to be just for me. But if you need someone to manage it for you, then there’s no privacy. (Workshop 1, Group 2, P5)}
\end{quote}

The desire to maintain personal autonomy sometimes led community members to exclude family members from certain healthcare activities, such as attending clinic visits alone, in order to preserve their independence and control:

\begin{quote}
    \textit{[My daughter said,] ``What’s your test result?'' I said, ``I don’t want to tell you.'' She said, ``I want to see your test result.'' I said, ``No.'' Then she said, ``Well, next time I’ll go with you.'' And I said, ``No, I’ll go by myself.'' (Workshop 2, Group 2, P24)}
\end{quote}

For P24, choosing to attend clinic visits alone was not just about privacy. It was a way to maintain autonomy and set boundaries to protect privacy, reducing dependence on younger generations even within supportive family relationships.

This subsection highlights how intergenerational relationships shape healthcare management in the Hmong community. Younger generations recognized the emotional and cultural value of the community, often through family relationships, while older generations received support with tasks such as scheduling appointments, using digital health platforms, and communicating with providers. Intergenerational support was viewed as an important asset, though some participants---especially older adults---expressed a desire to set clearer boundaries to maintain control over their health decisions. These examples show the need for careful approaches to maintaining privacy and agency while reinforcing community assets.

\subsection{Social Ties as an Asset for Accessing Quality Healthcare}

During the workshops, we learned how the strong social ties in the Hmong community became strengths, while also identifying some concerns, including worries about health information being shared informally or becoming widely known. These community-centered traditions and tight-knit relationships are assets that contribute to the community’s resilience in many ways, such that individuals can choose appropriate providers based on recommendations, share healthcare plans among family and friends, and build long-term relationships with providers. However, their close social ties also lead them to be cautious about their health information being shared within the community. As a result, they may avoid providers and interpreters from the same community, despite the need for culturally or linguistically appropriate care.

Participants frequently noted that extended kin networks often choose providers based on recommendations and share healthcare plans among their relatives and friends. They place high value on the opinions of those in their close networks, which influences their decisions about providers and treatment plans. P28 shared their observation of many community members choosing the same physician, noting that this allows the provider to become familiar with the community’s (clan’s) medical history and offer continuity of care:

\begin{quote}
    \textit{There's one particular doctor here in the clinic that a lot of Hmong patients like. It doesn’t matter how far they live---even if they have to drive for two hours, they still come to see this doctor. Sometimes, since the doctor knows the whole family, they’ll be like, ``Oh, how's your daughter? What is she doing nowadays?'' And then you’re more likely to feel comfortable opening up to that doctor. And again, she knows their symptoms and what they’ve been through with their health issues. (Workshop 2, Group 2, P28)}
\end{quote}

Similarly, others also shared that their extended family---including aunts, uncles, and cousins---all enrolled in the same health insurance plan. Seeing the same physician and being covered under the same insurance facilitates health management, as it allows extended families to share important practices (e.g., what to avoid to maintain healthy lives based on their lifestyles or genetic conditions), discuss possible treatment options, and make collective decisions that benefit both individuals and the broader community network.

Participants noted that their community members are not only engaged in healthcare practices within their close networks, but also strongly prefer long-term relationships with their healthcare providers, as they value human connection and trust in managing health. Participants expressed the benefits of continuing care, often staying with the same provider for decades despite logistical or financial challenges, such as clinic transfers, insurance limitations, or long travel distances. They indicated that these long-term relationships are assets and strengths of the community, as providers become familiar with each patient’s health history and adapt to their unique traits, such as personality and preferred communication style. One participant shared that over the past 45 years, they have only had five physicians:

\begin{quote}
    \textit{My first provider changed clinics, and my medical insurance wouldn’t cover the new one, so I had no choice. The second one, unfortunately, passed away. So I’ve only had four or five doctors since ’78. Yeah, I prefer it that way. And like I said, because we’ve been going to the same doctor, he knows exactly when I last saw him---that’s how I like it. (Workshop 2, Group 2, P22)}
\end{quote}

The participant shared how convenient it is to maintain the same provider, who is familiar with the details of the participant's care---even remembering things like the timing of past visits. Despite the importance of close social ties, participants emphasized that providers do not need to come from the same community. What matters more to them is having someone who genuinely understands their needs and provides long-term, consistent care. P22 further shared that they had never had a Hmong doctor, but it was not an issue for them.

Regardless of providers’ backgrounds, participants emphasized that a clear understanding of each patient’s history and the relationships is important in supporting their care. However, while strong social ties and close relationships bring benefits, community members should be cautious about their health information being shared within the community. They noted that in tight-knit communities like the Hmong community, helpful information often spreads quickly---but so do misinformation and personal details. P3 shared their frustrating experience of having the same provider as other community members:

\begin{quote}
    \textit{After my medical visit, many community and family members who see the same provider keep saying that if I take this medicine, it will make me sick, like this and like that. Then even if you stop, the side effects do not get better. There’s no follow-through when taking the medicine. And I get frustrated with that. (Workshop 1, Group 2, P3)}
\end{quote}

The participant mentioned that sharing the provider within the community often leads to unwanted advice about medications, which can create confusion and impact their treatment decisions. To mitigate potential negative consequences, participants described various efforts to safeguard their personal health information. These include seeking healthcare and choosing providers of other ethnicities (e.g., non-Hmong providers) and avoiding using specific services (i.e., interpreters) for health management. For example, female members of the Hmong community were often too embarrassed to disclose gynecological issues to their physicians. P1 mentioned that they prefer non-Hmong providers as these providers can alleviate their fears and concerns about being judged according to community norms:

\begin{quote}
    \textit{If I go to see the doctor and it’s something hard to hear or kind of embarrassing, then I want to see an American doctor, not a Hmong doctor. (Workshop 2, Group 1, P1)}
\end{quote}

Male participants shared similar sentiments, particularly regarding the discomfort of discussing private conditions such as prostate problems with female interpreters or family members:

\begin{quote}
    \textit{When it gets to the point of being embarrassed, I prefer to have a male provider so everyone feels more at ease. If it’s a family member, like daughters or daughters-in-law, I don’t say much. And if the interpreter is female, I feel too embarrassed. (Workshop 2, Group 1, P18)}
\end{quote}

For the same reason of protecting their health information, P11 shared their reluctance to involve interpreters, despite facing a language barrier and needing support. This participant mentioned feeling awkward about disclosing sensitive health issues in front of someone from the same community, due to the potential consequences of information spreading:

\begin{quote}
    \textit{The Hmong community is really small, and it’s very awkward when you get an interpreter and don’t want them to know your private information… Not everything might be shared, but some aspects of your situation are. The interpreters didn’t have training on how to maintain confidentiality. (Workshop 1, Group 2, P11)}
\end{quote}

As the quote illustrates, receiving necessary support without disclosing personal health information is difficult under current conditions, where there is limited training and awareness around protecting patient privacy in the healthcare environment.

This subsection illustrates how the strong social ties of the Hmong community shape health management practices. At the same time, participants expressed concerns about privacy, as their health information might circulate within the community. Their reflections highlight the importance of balancing the benefits of community support with the need to keep health information at the individual level.

\subsection{Storytelling as a Cultural Asset in Healthcare for Relatable and Personalized Care}

Another important asset highlighted in the workshops is the community’s strong storytelling tradition. The Hmong community has a longstanding oral tradition, with older adults playing a crucial role in passing down stories, including community history and the wisdom of previous generations. These linguistic and communication characteristics have become assets in Hmong patients’ health management practices. During the workshops, participants emphasized their preference for support that feels personal and relational, especially through in-person verbal interactions rather than written texts or digital information. Examples include storytelling and indirect communication styles that made participants feel understood and respected in culturally sensitive ways during consultations, conversational practices that provide opportunities for practical assistance tailored to individual needs, and appreciation for providers’ follow-up calls as expressions of care and commitment. Many community members also use voice-based platforms to access and share health information in ways that align with their communication strengths.

Because the Hmong language lacks many medical terms and expressions related to anatomy, clinical conversations often rely on storytelling techniques to explain symptoms and treatment options. Rather than explaining medical language directly, those involved in clinical communication adapt messages into forms that are culturally and linguistically resonant. One participant explained that Hmong community members often communicate medical terminology indirectly, by talking around the concept:

\begin{quote}
    \textit{If you translate the English version exactly, you’re out of place. Nothing will match. You have to go around it to make sure the other person understands. There are many English words we don’t even have in Hmong, and many Hmong words we don’t have in English. (Workshop 2, Group 2, P22)}
\end{quote}

Communication not only involves going around medical concepts to make them more understandable to Hmong patients, but also requires sensitivity to accommodate the unique characteristics of the Hmong language. For example, participants explained the importance of indirect communication methods, which avoid direct expressions such as ``no,'' ``failure,'' or ``death''---terms considered culturally inappropriate or insensitive. One participant from the second workshop shared specific examples of such expressions that are viewed as inappropriate within the community.

\begin{quote}
    \textit{In the Hmong community, there are just some words that you don't use. So I was listening to my grandson interpret the question, ``Have you ever thought of hurting yourself or killing yourself?'' He literally said in Hmong, ``Do you want to die?''---which is something we don't say. We say, ``Do you want to release your life?'' or ``Do you want to end your life?'' (Workshop 2, Group 3, P15)}
\end{quote}

As shown in this example, Hmong community members do not directly translate medical terminology. Instead, they carefully choose their words and tone, emphasizing the importance of adapting communication to reduce listeners’ fear and discomfort. This practice reflects the community’s deep cultural and linguistic roots in oral tradition.

Participants also shared community members’ positive experiences receiving assistance from providers in filling out health questionnaires and personal information, highlighting the value of conversational support in navigating healthcare processes (e.g., making clinic appointments).
Older Hmong patients particularly face challenges with text-heavy digital platforms and instead prefer verbal communication channels. They may call the clinic first, but if their calls end up in a queue, they often choose to visit the clinic in person and speak directly with staff at the check-in desk, rather than use digital methods such as email or patient portals. 

Similarly, their preference for verbal communication can be seen in the other regular healthcare processes, such as follow-up care. Participants valued follow-up phone calls from providers, which they viewed as meaningful expressions of care and connection.

\begin{quote}
    \textit{I have one doctor, and they really care about you. They say, ``I give you medication for a week, you come back and see how the medication helps you.'' After that, they follow up---``Is the medication helping you?'' They say, ``Okay, six weeks or six months, you come back, and I’ll take care of you and see how your health is.'' (Workshop 1, Group 1, P9)}
\end{quote}

As these quotes illustrate, participants emphasized during the workshop that follow-up phone calls provided a sense of care and comfort, which motivated them to follow their providers’ prescriptions.

Participants shared that their oral traditions strongly influence how they engage with technology-mediated platforms, expressing a preference for voice-centric tools like YouTube and radio to access and share health-related information in Hmong. One participant shared how their family listened to Hmong content on YouTube; regardless of age, these social media platforms allowed them to enjoy the community’s storytelling culture and pass down meaningful stories.

\begin{quote}
    \textit{We know YouTube because we have been watching Hmong videos, Hmong news, and Hmong entertainment every day. I am also doing a YouTube channel myself. (Workshop 1, Group 3, P10)}
\end{quote}

As P10 noted, community members tend to engage with health information in ways that align with their preference for verbal communication, particularly when it is delivered in familiar forms. Other examples mentioned include group video chats, which allow for more personal and interactive communication.

This subsection shows how the Hmong community’s preference for verbal communication and their storytelling practices shape the pre-visit, during-visit, and post-visit stages of healthcare management practices. Participants preferred verbal communication throughout their healthcare journey, and emphasized the importance of culturally appropriate communications, including the use of indirect language.


\section{Discussion}

From the workshops, our participants’ experiences in balancing their traditional healthcare practices within their new environment highlight design opportunities for healthcare technologies that support newcomers as they navigate and build their lives in unfamiliar settings (RQ1). We also show how technologies can be designed to accommodate and respect traditional practices, supporting intergenerational and community-based health practices while minimizing concerns around privacy and autonomy (RQ2).

\subsection{Design Opportunities to Support Comfort and Confidence in Cultural Health Practices}

Through our participatory design workshops, we learned how community members navigate healthcare by combining traditional practices---such as herbal medicine, consultations with shamans, and cultural rituals---with care provided by their doctors. These traditional practices offer emotional comfort, social connection, and family-based care that help them remain resilient in their new environment. As one young participant noted, even after several decades of resettlement, these practices continue to serve as a way to feel care and support from family and community.

These findings align with prior studies in HCI and CSCW, which have shown the potential for technologies to support spiritual and reflective practices during significant life challenges (e.g., health conditions~\cite{smith_etal_WhatSpiritualSupport_2021} and resettling in a new environment~\cite{almohamed_vyas_RebuildingSocialCapital_2019}). These studies discussed how spirituality, faith, and religious practices can be supported through technology-mediated platforms, including online communities and mobile applications~\cite{rosner_etal_SpiritualityDesign_2022, prinster_etal_CommunityArchetypesEmpirical_2024}. Engagement in those practices through various platforms helps individuals foster a sense of belonging, remain grounded in their beliefs, and maintain their cultural identity---providing resilience and strength to navigate the challenges they face in their daily lives~\cite{ibrahim_etal_TrackingRamadanExamining_2024}.

In our workshops, participants shared how they balance and selectively adopt traditional forms of care and Western medicine for different purposes. They see these practices as complementary rather than mutually exclusive. The ability to rely on traditional practices---which offer a sense of belonging, care, and comfort---alongside receiving care from doctors in clinical settings is a resource and strength that participants have developed over time through lived experiences of navigating healthcare in a new environment. However, at the same time, participants shared feelings of guilt or stigma associated with using traditional care practices, which may hinder their ability to fully benefit from these forms of care. Although our participants draw emotional comfort, social connection, and a sense of healing from their cultural practices (e.g., herbal medicine and prayer), they also expressed fear of being judged because such practices are often viewed as informal, less legitimate, or non-scientific within mainstream healthcare, making it difficult to openly discuss the practices outside of their communities.

Previous studies also showed that community members felt a sense of guilt in managing intimate health information and were reluctant to talk to healthcare providers during religiously-significant times~\cite{al-naimi_alistar_UnderstandingCulturalReligious_2024, stowell_etal_InvestigatingOpportunitiesCrowdsourcing_2020}. In response to these challenges, this line of research has proposed design strategies that respect and prioritize religious norms and traditions when supporting individual health management, such as personal health data tracking~\cite{mustafa_etal_PatriarchyMaternalHealth_2020}. Tailoring language and contextual awareness (e.g., the dates of religious events) were suggested as useful design implications. 
For example, studies proposed delivering health information within culturally tailored frameworks by adapting language to avoid secular or overly clinical terms~\cite{ibrahim_etal_TrackingRamadanExamining_2024}. This included connecting healthcare practices to community responsibilities or using indirect wording to reduce unnecessary tension---for instance, replacing ``sexual health'' with ``women’s health''~\cite{al-naimi_alistar_UnderstandingCulturalReligious_2024}. Another study emphasized the importance of timing when communicating health information during culturally or religiously significant periods, particularly leveraging asynchronous methods to show cultural sensitivity~\cite{al-naimi_alistar_UnderstandingCulturalReligious_2024}. For instance, one study suggested waiting until after prayer events had concluded to broadcast health information on the community’s social media platforms, to respect the sacred space while still reaching the community~\cite{stowell_etal_InvestigatingOpportunitiesCrowdsourcing_2020}.

We propose that technology-mediated solutions can support Hmong patients' cultural practices more effectively and safely. These solutions can help reduce unnecessary feelings of guilt, promote openness and confidence in managing health, and support patients in balancing their cultural practices with formal healthcare system. During the workshops, participants shared their familiarity with verbal-oriented social media platforms to receive and share information, which they found to be more relatable and useful. Building on this finding, we suggest expanding the application of these platforms for healthcare use. For example, voice notes, podcasts, or interactive applications can be used to deliver health information in culturally resonant ways, such as through storytelling or group forums. These platforms could feature trusted community leaders or healthcare providers from the Hmong community as a key design consideration, delivering important health messages or sharing personal stories about healthcare practices in culturally appropriate and relatable tones. Engaging with sensitive topics through familiar and trusted communication styles may help community members feel less guilt or hesitation, and view their cultural practices not as informal or stigmatized, but as meaningful resources integrated into their health management. Importantly, delivering this content through reliable and appropriately screened media channels, such as trusted social media groups or clinic-affiliated messaging apps, can help ensure that the information is accurate, safe, and culturally sensitive. These approaches may also help reduce the sense of taboo or embarrassment often associated with discussing traditional care practices. In addition, these resources could also serve as learning tools for healthcare providers, helping them better understand small or underrepresented communities and suggest more culturally competent care practices. By fostering a sense of legitimacy, trust, and safety, these strategies can support patients in balancing their cultural practices with clinical care, enabling them to find comfort in their traditions while receiving appropriate guidance and support from the healthcare system.

\subsection{Design Opportunities for Managing Privacy Boundaries in Collaborative Health Practices}

Previous studies on healthcare management within domestic settings have identified how families effectively collaborate in managing health and emphasized the benefits of tracking health information~\cite{cha_etal_CollaborativeHealthtrackingTechnologies_2025, cha_etal_SharedResponsibilityCollaborative_2024, shin_holtz_BetterTransitionsChildren_2019, maestre_etal_NotAnotherMedication_2021}, receiving emotional support~\cite{nikkhah_etal_DesigningFamiliesAdaptive_2022}, and preserving family rituals~\cite{shin_etal_MoreBedtimeBedroom_2022, shin_etal_BedtimePalsDeployment_2023}. For example, Pina et al. identified the critical mediating roles of younger family members in assisting their parents with limited English proficiency to access and interpret diagnoses and treatment plans after clinic visits~\cite{pina_etal_HowLatinoChildren_2018}. Similarly, another study highlighted how collective decision-making around healthy eating, supported by strong social ties within the family, can effectively facilitate behavior change through mutual influence among family members in low-resource communities~\cite{schaefbauer_etal_SnackBuddySupporting_2015}. 

Echoing these previous studies, our findings show that the Hmong community’s strong social networks---particularly with family and relatives---are deeply embedded in their healthcare practices. Younger family members often assist older generations with healthcare technologies, such as creating and managing patient portals, while also learning traditional knowledge and wisdom from elders. These close intergenerational relationships influence how community members choose healthcare providers and make treatment decisions. This strong social connectedness represents a key asset of the Hmong community, which could be further leveraged to enhance their everyday health management practices. At the same time, however, these close social networks can introduce unique challenges around privacy and autonomy. Privacy concerns are particularly important in vulnerable populations because managing health often happens within close family or community networks~\cite{ghaiumyanaraky_etal_DiscloseNotDisclose_2021, schaefbauer_etal_SnackBuddySupporting_2015}. Since sharing sensitive information (e.g., urological or gynecological issues) is not only a personal decision, but one that can affect relationships, reputation, or trust within the community, individuals may feel additional pressure or discomfort in managing health information, especially when relying on others for support. Consistent with this, our findings showed that privacy tensions arise within intergenerational and community-centered health management for Hmong individuals---even while they recognize the need for family or community support. Similar concerns applied to the use of interpreters from their own community, raising fears of potential breaches of confidentiality. 

HCI and CSCW studies have emphasized related challenges around shared communication devices and the discomfort of disclosing sensitive information, especially among vulnerable populations (e.g., immigrant women~\cite{brown_etal_ReflectionDesignImmigrant_2014} and left-behind children~\cite{wong-villacres_bardzell_TechnologymediatedParentchildIntimacy_2011}). These concerns sometimes led participants (particularly older adults) to avoid specific technologies or healthcare services to maintain privacy. Recent studies have explored privacy issues in managing data and information within close relationships (e.g., between parents and adult children, or in families with teenagers) across various contexts, including healthcare~\cite{li_etal_PrivacyVsAwareness_2023, wang_etal_FamiDataHubSpeculative_2025, shin_holtz_BetterTransitionsChildren_2019}. These studies highlight the trade-offs between maintaining awareness of family situations and respecting individual privacy. They also discuss different strategies, such as task- or content-specific privacy management and clear communication about what information is being shared. 

We suggest that concerns around privacy and autonomy should not prevent or discourage intergenerational and community-based collaboration, which are critical assets in health management for resettled communities. To support these practices, technologies should provide users with flexible control over boundaries, including decisions about when, what, and with whom health information is shared, to help them manage disclosure in ways that feel safe and appropriate. Several studies have applied family-centered perspectives to examine the role of family members in managing health, particularly within resettled communities. For instance, one study showed that Latino parents often rely on their children to search for health-related information, to compensate for their own limited digital literacy---though they sometimes hesitate to share sensitive contextual details, such as a sibling’s chronic health conditions, due to fears of emotionally burdening their child~\cite{pina_etal_HowLatinoChildren_2018}. However, there remains a notable gap in the healthcare domain regarding how technologies could effectively support complex family dynamics and situational needs over time, particularly in ways that help patients manage privacy, control information boundaries, and maintain autonomy beyond parent--young child relationships.

To alleviate concerns around privacy in family-centered health management while maximizing the benefits of intergenerational and community support, technology-mediated health management platforms (particularly patient portals) could incorporate context-sensitive or time-delayed information sharing features. These features would enable patients to control both who can access their health information and when it is shared, depending on the sensitivity of the content (e.g., urological or gynecological issues identified in our study) and the nature of family relationships involved in care (e.g., male patient--daughter-in-law relationships).

However, as participants in our workshops explained, patients often depend on younger family members to create and manage accounts for health technology platforms, which suggests that this feature alone may not be sufficient. Careful design approaches are needed, such as granting temporary access modes where family members assist with setup or technical tasks but cannot view actual content unless given time-specific permission. Verbal or audio-based delivery of less sensitive information through family or friends could also leverage close social ties to facilitate access to necessary health information while keeping other content private. Such patient-centered designs that consider family involvement would enable community members to benefit from shared care practices while preserving patient dignity, privacy, and autonomy---even in situations where technology adoption is supported by others within their close social networks. These features offer culturally appropriate ways of managing sensitive information within tightly connected families and communities, while still maintaining the supportive role of family in care.
With these design approaches, healthcare systems could respect community members’ autonomy while fostering family-centered care. They would also provide patients with time and space to process important medical information, and help reduce social pressure when navigating sensitive health topics.

\section{Limitations and Future Work}
In our community-based workshops, we included participants from one ethnic community, the Hmong, and proposed design implications based on their strengths and cultural characteristics. Although these implications are tailored for the Hmong community, they could be applied to and tested with other similar small communities, considering the common challenges and strengths in resettled communities. As the next step, hands-on design activities (i.e., co-design) with community members could involve prototyping different technology-mediated solutions, such as social robots addressing privacy boundaries and techno-spirituality platforms. Additionally, during our data collection and analysis, we identified herbal medicine, shamanism, and spirituality as important topics that deserve more focused attention from researchers in the field. Future studies should investigate how technology-mediated approaches can engage with these unique shamanic practices and explore their potential integration into healthcare, ultimately contributing to improved health outcomes.

\section{Conclusion}
Through two community-based workshops, we identified four types of assets that community members developed in managing their health. While previous studies proposed various technology-mediated solutions to mitigate the community’s deficits and challenges, we focused on the cultural strengths of the community to address our research questions. The proposed interventions could support community members' comfort and confidence in cultural practices while balancing them with providers' prescriptions. We also suggested flexible ways to have control over boundaries, including decisions about when, what, and with whom health information is shared, to help community members manage disclosure in ways that feel safe and appropriate. Our discussion underscores the importance of culturally tailored, community-based participatory design approaches grounded in the lived experiences of resettled communities. By illustrating how cultural assets and needs influence the development of health technologies, this work offers valuable insights into more effective, community-driven solutions. While centered on the Hmong community, these contributions may also inform the design of health management technologies for other resettled populations facing similar challenges in navigating unfamiliar healthcare systems. 

\begin{acks}
 This work was supported by the University of Minnesota’s Social Justice Impact Grant. We thank Adam Thao for his contributions during the workshop. We also thank the community members who participated in the workshops and shared their valuable experiences with us. 
\end{acks}

\bibliographystyle{ACM-Reference-Format}
\bibliography{sample-base}


\end{document}